\documentclass[conference]{IEEEtran}
\ifCLASSINFOpdf
\else
\fi
\usepackage{citesort,color,graphicx,amssymb,amsmath,commath,amsthm}
\usepackage{caption}
\usepackage{comment}

\IEEEoverridecommandlockouts
\newtheorem{claim}{Claim}
\graphicspath{{./} {figures/}}
\begin{document}
\title{
Mismatched Estimation in Large Linear Systems}
\author{{\bf Yanting Ma,$^{\dagger}$ Dror Baron,$^\dagger$ Ahmad Beirami$^*$}\vspace{0.05in}\\
$^\dagger$Department of Electrical and Computer Engineering, North Carolina State University, Raleigh, NC 27695, USA\\
$^*$Department of Electrical and Computer Engineering, Duke University, Durham, NC 27708, USA\vspace{0.05in}\\
Email: $^\dagger$\{yma7, barondror\}@ncsu.edu, $^*$ahmad.beirami@duke.edu
\thanks{This work was supported in part by the National Science Foundation under grant CCF-1217749 and in part by the U.S. Army Research Office under grant W911NF-14-1-0314.}
}
\maketitle
\thispagestyle{empty}

\begin{abstract}
We study the excess mean square error (EMSE) 
above the minimum mean square error (MMSE) in large linear systems
where the posterior mean estimator (PME) is evaluated with a 
postulated prior that differs from the true prior of the input signal. 
We focus on large linear systems where the measurements are acquired via an independent and identically distributed random matrix, and are corrupted by additive white Gaussian noise (AWGN).
The relationship between the EMSE in large linear systems and EMSE in scalar channels is derived, and closed form approximations are provided. 
Our analysis is based on the
decoupling principle, which links scalar channels to large linear system analyses. 
Numerical examples demonstrate that our closed form approximations are accurate.
\end{abstract}
\begin{IEEEkeywords}
decoupling,
large linear systems,  
mismatched estimation.
\end{IEEEkeywords}

\section{Introduction}
\label{sec:intro}

The posterior mean estimator (PME), 
also known as conditional expectation, 
plays a pivotal role in Bayesian estimation. 
To compute the PME, we need a prior distribution for the unknown signal. 
In cases where the prior is unavailable, we may compute the PME with a 
postulated prior, which may not match the true prior. 
Verd{\'u}~\cite{Verdu2010} studied the mismatched estimation problem 
for scalar additive white Gaussian noise (AWGN) channels
and quantified the excess mean square error (EMSE) above the minimum mean 
square error (MMSE) due to the incorrect prior.
A natural extension to Verd{\'u}'s result would be to quantify the EMSE due to mismatched estimation in large linear systems~\cite{Tanaka2002,GuoVerdu2005,GuoWang2007,GuoBaronShamai2009,DMM2009,RFG2012}.

\vspace{0.08in}
{\bf Mismatched estimation:}
Consider scalar estimation,
\begin{equation}
Y=X+\sigma W,\label{eq:scalar_channel}
\end{equation}
where $X$ is generated by some probability density function (pdf) $p_X$, $W\sim\mathcal{N}(0,1)$ is independent of $X$, and $\mathcal{N}(\mu,\sigma^2)$ denotes the Gaussian pdf with mean $\mu$ and variance $\sigma^2$.
A PME with some prior $q_X$, which is defined as
\begin{equation}
\label{eq:mismatched estimator}
\widehat{X}_q(y;\sigma^2)=\mathbb{E}_{q_X}[X|Y=y],
\end{equation}
can be applied to the estimation procedure,
where in $\mathbb{E}_{q_X}[\cdot]$ expectation is calculated assuming that $X$ is distributed as $q_X$.
The mean square error (MSE) achieved by $\widehat{X}_q(y;\sigma^2)$ is
\begin{equation}
\Psi_q(\sigma^2)=\mathbb{E}_{p_X}\left[(\widehat{X}_q(\cdot;\sigma^2)-X)^2\right].
\label{eq:MSE}
\end{equation}
Note that $\widehat{X}_p(y;\sigma^2)$ is the MMSE estimator, and $\Psi_p(\sigma^2)$ is the MMSE.

Having defined notation for the MSE, we can now define the EMSE above the MMSE due to the mismatched prior in the scalar estimation problem~(\ref{eq:scalar_channel}):
\begin{equation}
\text{EMSE}_s(\sigma^2)=\Psi_q(\sigma^2)-\Psi_p(\sigma^2),\label{eq:EMSE_s_def}
\end{equation}
where the subscript $s$ represents scalar estimation.

Verd{\'u}~\cite{Verdu2010} proved that $\text{EMSE}_s$ 
is related to the relative entropy~\cite{Cover06} as follows:
\begin{equation*}
\text{EMSE}_s(\sigma^2)  = \frac{\textsf{d}}{\textsf{d}\gamma} D\left(P_X\star\mathcal{N}\left(0,\frac{1}{\gamma}\right)
\right\|\left. Q_X\star\mathcal{N}\left(0,\frac{1}{\gamma}\right)\right),
\end{equation*}
where $\star$ represents convolution, $\gamma=1/\sigma^2$, $P_X$ and $Q_X$ are the true and mismatched probability distributions, respectively, and under the assumption that $P$ is absolutely continuous with respect to $Q$, the relative entropy, also known as Kullback-Leibler divergence~\cite{Cover06}, is defined as $D(P\|Q)=\int \log\left(\frac{\textsf{d}P}{\textsf{d}Q}\right)\textsf{d}P$.

\vspace{0.08in}
{\bf Large linear system:}                                                     
Consider a linear system:
\begin{equation}
{\bf y} = {\bf Ax} + {\bf z},\label{eq:matrix_channel}
\end{equation}
where the input signal ${\bf x}\in\mathbb{R}^N$ is a sequence of independent and identically distributed (i.i.d.)
random variables generated by some pdf $p_X$, ${\bf A}\in\mathbb{R}^{M\times N}$ is an i.i.d. Gaussian random matrix with $A_{ij}\sim\mathcal{N}(0,1/\sqrt{M})$, and ${\bf z}\in\mathbb{R}^M$ is AWGN with mean zero and variance $\sigma_z^2$. Large linear systems~\cite{Tanaka2002,GuoVerdu2005,GuoWang2007,GuoBaronShamai2009,DMM2009,RFG2012}, which are sometimes called the large system limit of linear systems,
refer to the limit that both $N$ and $M$ tend to infinity but with their ratio converging to a positive number, i.e., $N\rightarrow\infty$ and $M/N\rightarrow\delta$. We call $\delta$ the measurement rate of the linear system.

\vspace{0.08in}
{\bf Decoupling principle:}
The decoupling principle~\cite{GuoVerdu2005}
is based on replica analysis of randomly spread code-division 
multiple access (CDMA) and multiuser detection. 
It claims the following result. 
Define the PME with prior $q_X$ of the linear system~(\ref{eq:matrix_channel}) as
\begin{equation}
\widehat{{\bf x}}_q=\mathbb{E}_{q_X}[{\bf x}|{\bf y}].\label{eq:PME_vec}
\end{equation}
Denote the $j$-th element of the sequence ${\bf x}$ by $x(j)$.
For any $j\in\{1,2,...,N\}$, the joint distribution of $(x(j),\widehat{x}_q(j))$
converges in probability to the joint distribution of $(X,\widehat{X}_q(\cdot;\sigma_q^2))$
in large linear systems, 
where $X\sim p_X$ and $\sigma_q^2$ is the solution to 
the following fixed point equation~\cite{Tanaka2002,GuoVerdu2005}, 
\begin{equation}
\delta\cdot(\sigma^2-\sigma^2_z)=\Psi_q(\sigma^2).\label{eq:SE}
\end{equation}
Note that
$\widehat{X}_q(\cdot;\sigma_q^2)$ is defined in (\ref{eq:mismatched estimator}), 
and there may be multiple fixed 
points~\cite{GuoVerdu2005,Krzakala2012probabilistic,ZhuBaronCISS2013}. 

The decoupling principle provides a single letter characterization of 
the MSE; 
state evolution~\cite{Bayati2011} in approximate message passing 
(AMP)~\cite{DMM2009} provides rigorous justification for the 
achievable part of this characterization.
An extension of the decoupling principle 
to a collection of any finite number of elements in ${\bf x}$ 
is provided by Guo et al.~\cite{GuoBaronShamai2009}.

\begin{figure}[t!]
\center
\includegraphics[width=\linewidth]{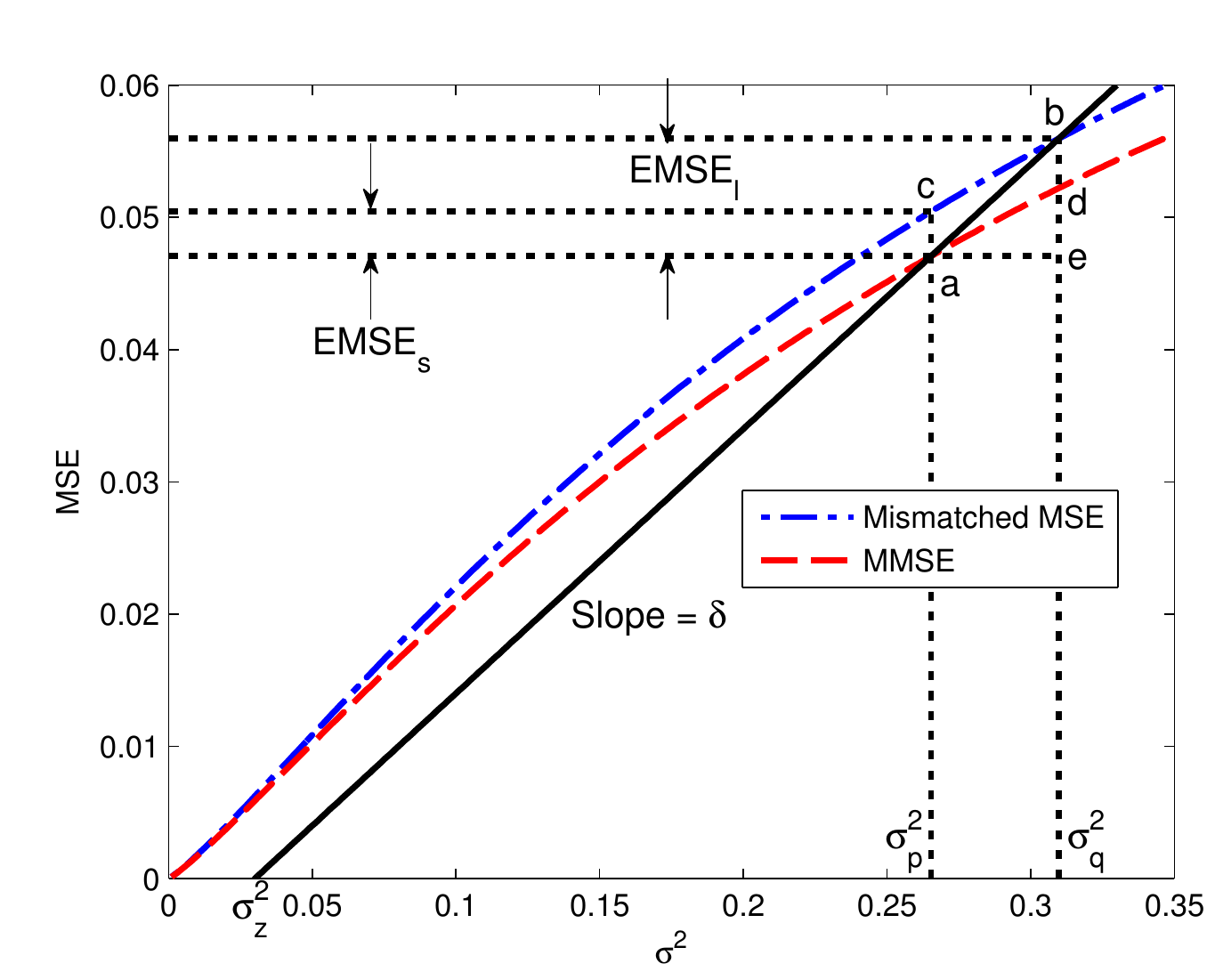}
\caption{Mismatched estimation for Bernoulli-Gaussian input signal 
($p_X(x)=(1-\theta)\delta_0(x)+\theta\mathcal{N}(0,1))$
in large linear systems. 
We notice that the EMSE in the scalar estimation problem is {\em amplified} 
in the large linear system, despite using the same mismatched prior for PME,
because the decoupled scalar channel becomes noisier
as indicated by the horizontal move from c to b. 
($\delta=0.2$, $\sigma_z^2=0.03$, the true sparsity parameter $\theta=0.1$, 
and the mismatched sparsity parameter $\widetilde{\theta}=0.2$.)}
\label{fig:AmpEffect}
\end{figure}

\vspace{0.08in}
{\bf Illustration:}\
Figure~\ref{fig:AmpEffect} highlights our contribution, which is stated in Claim~\ref{claim:main},
using an example that compares the EMSE in scalar channels 
and large linear systems. 
The solid line with slope $\delta$ represents the linear function of $\sigma^2$ 
on the left-hand-side of (\ref{eq:SE}). 
The dashed and dash-dotted curves represent $\Psi_p(\sigma^2)$ and $\Psi_q(\sigma^2)$, respectively. 
The intersection point $a$ provides the solution to the fixed point equation (\ref{eq:SE}) when $\widehat{X}_p$ is applied, and so the Cartesian coordinate representation of $a$ is $a=(\sigma_p^2,\Psi_p(\sigma_p^2))$. Similarly, we have $b=(\sigma_q^2,\Psi_q(\sigma_q^2))$. 
Therefore, the vertical distance between $a$ and $b$ is the difference 
in MSE of the two decoupled channels for the PME with the true prior and the 
mismatched prior, respectively.
Based on the decoupling principle, the vertical distance is equivalent to 
the EMSE in large linear systems, which we denote by EMSE$_l$ and define as 
\begin{equation}
\text{EMSE}_l
=\lim_{N\rightarrow\infty}\frac{1}{N}\mathbb{E}_{p_X}\left[\|\widehat{\bf x}_q-{\bf x}\|_2^2-\|\widehat{\bf x}_p-{\bf x}\|_2^2\right],
\label{eq:EMSE_l_def_org}
\end{equation}
where the subscript $l$ represents large linear systems, 
$\widehat{\bf x}_q$ is defined in (\ref{eq:PME_vec}), 
and $\widehat{\bf x}_p$ is defined similarly.
The pair $c=(\sigma_p^2,\Psi_q(\sigma_p^2))$ represents the MSE achieved by 
$\widehat{X}_q$ in the decoupled scalar channel for $\widehat{X}_p$, 
and so the vertical distance between $a$ and $c$ is $\text{EMSE}_s(\sigma_p^2)$.
We can see from Figure~\ref{fig:AmpEffect} that $\text{EMSE}_l$ is larger than $\text{EMSE}_s(\sigma_p^2)$, despite using the same mismatched prior for PME,
because the decoupled scalar channel becomes noisier
as indicated by the horizontal move from c to b; we call this an amplification effect.
{\em Our contribution is to formalize the amplification of $\text{EMSE}_s(\sigma_p^2)$ to $\text{EMSE}_l$.}  

The remainder of the paper is organized as follows. 
We derive the relationship between 
the EMSE in large linear systems and EMSE in scalar channels in 
Section~\ref{sec:mismatch}; 
closed form approximations that characterize this relationship are also provided. 
Numerical examples that evaluate the accuracy level of our closed form approximations are presented in Section~\ref{sec:numerical}, and Section~\ref{sec:conclusion} concludes the paper.

\section{Main Results}
\label{sec:mismatch}

We now characterize the relationship between $\text{EMSE}_l$ and 
$\text{EMSE}_s(\sigma_p^2)$, which is the EMSE of the mismatched PME in the 
decoupled scalar AWGN  channel for the PME with the true prior $p_X$.

Our main result is summarized in the following claim. We call this result a claim,
because it relies on the decoupling principle~\cite{GuoVerdu2005}, 
which is based on the replica method and lacks rigorous justification.

\begin{claim} \label{claim:main}
Consider a large linear system~(\ref{eq:matrix_channel}). 
Let the EMSE in scalar channels $\text{EMSE}_s(\sigma^2)$ and 
EMSE in large linear systems $\text{EMSE}_l$ be defined in 
(\ref{eq:EMSE_s_def}) and (\ref{eq:EMSE_l_def_org}), respectively, 
and let $\sigma_p^2$ be the noise variance in the decoupled 
scalar channel when the true prior $p_X$ is applied. Denote $\Psi_q'(\sigma^2)=\frac{\textsf{d}}{\textsf{d}(\sigma^2)}\Psi_q(\sigma^2)$, where $\Psi_q(\sigma^2)$ is the MSE in the scalar AWGN channel with noise variance $\sigma^2$ achieved by the PME with a mismatched prior $q_X$~(\ref{eq:MSE}). In large linear systems, $\text{EMSE}_l$ and $\text{EMSE}_s(\sigma_p^2)$ satisfy the following relation:
\begin{equation}
\text{EMSE}_l=\text{EMSE}_s(\sigma_p^2)+\int_{\sigma_p^2}^{\sigma_p^2+\frac{1}{\delta}\text{EMSE}_l}\Psi_q'(\sigma^2)\textsf{d}\sigma^2.
\label{eq:claim}
\end{equation}
\end{claim}

{\em Justification:}
The fixed point equations~(\ref{eq:SE}) when applying $\widehat{X}_p$ and $\widehat{X}_q$ are
\begin{align}
\delta\cdot\left(\sigma_p^2-\sigma_z^2\right)&=\Psi_p(\sigma_p^2) \quad\text{and}\label{eq:fix_pt_p}\\
\delta\cdot\left(\sigma_q^2-\sigma_z^2\right)&=\Psi_q(\sigma_q^2),\label{eq:fix_pt_q}
\end{align}
respectively. 
Subtract (\ref{eq:fix_pt_p}) from (\ref{eq:fix_pt_q}):
\begin{equation}
\delta\cdot\left(\sigma_q^2-\sigma_p^2\right)=\Psi_q(\sigma_q^2)-\Psi_p(\sigma_p^2)\label{eq:delta_sigma2}.
\end{equation}
Recall that
\begin{align}
\text{EMSE}_l &=\Psi_q(\sigma_q^2)-\Psi_p(\sigma_p^2),\label{eq:EMSE_l_j}\\
\text{EMSE}_s(\sigma_p^2)&=\Psi_q(\sigma_p^2)-\Psi_p(\sigma_p^2).\nonumber
\end{align}
Combining the two equations:
\begin{align}
\text{EMSE}_l&=\text{EMSE}_s(\sigma_p^2)+\Psi_q(\sigma_q^2)-\Psi_q(\sigma_p^2)\label{eq:claim_no_int}\\
&=\text{EMSE}_s(\sigma_p^2)+\int_{\sigma_p^2}^{\sigma_p^2+\left(\sigma_q^2-\sigma_p^2\right)}\Psi_q'(\sigma^2)\textsf{d}\sigma^2,\nonumber
\end{align}
where (\ref{eq:claim}) follows by noticing from (\ref{eq:delta_sigma2}) and (\ref{eq:EMSE_l_j}) that $\sigma_q^2-\sigma_p^2=\frac{1}{\delta}\text{EMSE}_l$.\qed

\vspace{0.08in}
{\bf Approximations:} Consider a Taylor expansion of $\Psi_q(\sigma_q^2)$ at $\sigma_p^2$:
\begin{align}
\Psi_q(\sigma_q^2) &= \Psi_q(\sigma_p^2)+\alpha(\sigma_q^2-\sigma_p^2)+\frac{1}{2}\beta(\sigma_q^2-\sigma_p^2)^2\nonumber\\
&+o((\sigma_q^2-\sigma_p^2)^2)\nonumber\\
&= \Psi_q(\sigma_p^2)+\alpha\frac{\text{EMSE}_l}{\delta}+\frac{\beta}{2}\left(\frac{\text{EMSE}_l}{\delta}\right)^2\nonumber\\
&+o\left(\Delta^2\right)\label{eq:Taylor},
\end{align}
where $\alpha=\Psi_q'(\sigma_p^2)$ and $\beta=\Psi_q''(\sigma_p^2)$ are the first and second derivatives, respectively, of $\Psi_q(\cdot)$ evaluated at $\sigma_p^2$, and $\Delta=\sigma_q^2-\sigma_p^2$.\footnote{$h(\Delta)=o(g(\Delta))$ if $\displaystyle\lim_{\Delta\rightarrow 0}\displaystyle\frac{h(\Delta)}{g(\Delta)}=0$.}

Plug (\ref{eq:Taylor}) into (\ref{eq:claim_no_int}):
\begin{equation}
\text{EMSE}_l=\text{EMSE}_s(\sigma_p^2)+\alpha\frac{\text{EMSE}_l}{\delta}+\frac{\beta}{2}\left(\frac{\text{EMSE}_l}{\delta}\right)^2+o\left(\Delta^2\right).\label{eq:solving_for_EMSE_l}
\end{equation}
If we only keep the first order terms in (\ref{eq:solving_for_EMSE_l}) and solve for $\text{EMSE}_l$,
then the first order approximation of $\text{EMSE}_l$ is obtained in closed form:
\begin{equation}
\text{EMSE}_l =
\frac{\delta}{\delta-\alpha}
\text{EMSE}_s(\sigma_p^2)
+o\left(\Delta\right).
\label{eq:appr_1st}
\end{equation}
Note that $ \frac{\delta}{\delta-\alpha}>1$. That is, the EMSE in the scalar 
estimation problem is {\em amplified} in large linear systems despite 
using the same mismatched prior $q_X$ for PME. This is due to an increased 
noise variance in the decoupled channel for the mismatched prior beyond the 
variance for the correct prior.

Similarly, the second order approximation is given by
\begin{align}
\text{EMSE}_l&=\frac{\delta\text{EMSE}_s(\sigma_p^2)}{\delta-\alpha}\left(1+\frac{1}{2}\frac{\beta\text{EMSE}_s(\sigma_p^2)}{(\delta-\alpha)^2}\right)+o\left(\Delta^2\right),\label{eq:appr_2nd_appr}
\end{align}
under the condition that $\Psi_q(\sigma^2)$ is locally concave in $\sigma^2$ 
around $\sigma_p^2$; details in the Appendix.  A more accurate approximation is also presented in the Appendix in~\eqref{eq:appr_2nd}.

We expect that when the mismatched distribution $q_X$ is close to $p_X$, $\Psi_p(\sigma^2)$ and $\Psi_q(\sigma^2)$ are close to each other for all $\sigma^2$, and $\Delta=\sigma_q^2-\sigma_p^2$ is smaller for minor mismatch than significant mismatch with the same slope of $\Psi_q(\sigma^2)$ at $\sigma_p^2$ and the same $\delta$. Therefore, the first order approximation of $\Psi_q(\sigma^2)$ for $\sigma^2\in[\sigma_p^2,\sigma_p^2+\Delta]$ is more likely to be reasonably accurate for minor mismatch. When the mismatch is significant, we might need to include the second order term in the Taylor expansion~(\ref{eq:Taylor}) to
improve accuracy. Numerical examples that show the necessity of the second order 
term when there is significant mismatch will be shown in Section~\ref{sec:numerical}.

\begin{table}[t!]
\caption{Relative error in the Bernoulli example} 
\centering 
\begin{tabular}{c c c c c} 
\hline
$\widetilde{\theta}$ & $\Delta$ & 1st (\ref{eq:appr_1st}) &2nd (\ref{eq:appr_2nd_appr}) &2nd (\ref{eq:appr_2nd})\\ 
\hline 
0.11 & 0.0008 & 0.13\%  &$<$0.0005\%  & $<$0.0001\%\\
0.13 & 0.0070 & 1\%     &0.041\%      & 0.017\%  \\
0.15 & 0.0178 & 2.8\%   &0.28\%       & 0.11\%\\  
0.17 & 0.0324 & 5.2\%   &0.99\%       & 0.35\%\\ 
0.20 & 0.0603 & 10\%    &4\%          & 1\%\\ 
\hline 
\end{tabular}
\label{table:Bernoulli} 
\end{table}

\section{Numerical Examples}
\label{sec:numerical}

\subsection{Accuracy of approximations}

We begin with two examples that examine the accuracy level of 
our first and second order approximations given by~(\ref{eq:appr_1st})-(\ref{eq:appr_2nd}). 
 
 \vspace{0.08in}
{\bf Example 1: Bernoulli input.}
The input signal of the first example follows a Bernoulli distribution with 
$p_X(1)=\theta$ and $p_X(0)=1-\theta$. Let $\theta=0.1$, and let the mismatched Bernoulli parameter $\widetilde{\theta}$ vary from 0.11 (minor mismatch) to 0.20 (significant mismatch). 
The linear system~(\ref{eq:matrix_channel}) has measurement rate 
$\delta=0.2$ and measurement noise variance $\sigma_z^2=0.03$. 
Using (\ref{eq:SE}), this linear system with the true Bernoulli prior
yields $\sigma_p^2=0.34$.
Table~\ref{table:Bernoulli} shows the accuracy level of our three approximations~(\ref{eq:appr_1st}), (\ref{eq:appr_2nd_appr}) and~(\ref{eq:appr_2nd}) 
for the Bernoulli example. The relative error of the predicted $\text{EMSE}_l$, 
which is denoted by $\text{EMSE}_l(\text{pred})$, 
is defined as $|\text{EMSE}_l-\text{EMSE}_l(\text{pred})|/\text{EMSE}_l$, 
where the first and second order approximations for $\text{EMSE}_l(\text{pred})$
are given by (\ref{eq:appr_1st})-(\ref{eq:appr_2nd}), respectively. 

 \vspace{0.08in}
{\bf Example 2: Bernoulli-Gaussian input.}
Here the input signal follows a Bernoulli-Gaussian distribution 
$p_X(x) = \theta\mathcal{N}(0,1)+(1-\theta)\delta_0(x)$, 
where $\delta_0(\cdot)$ is the delta function~\cite{Papoulis91}. 
Let $\theta=0.1$, and let the mismatched parameter 
$\widetilde{\theta}$ vary from 0.11 to 0.20 as before. 
The linear system is the same as in the Bernoulli example
with $\delta=0.2$ and $\sigma_z^2=0.03$, 
and this linear system with the correct Bernoulli-Gaussian prior 
yields $\sigma_p^2=0.27$. 
Figure~\ref{fig:AmpEffect} compares the PME with the true parameter 
and the PME with the mismatched parameter $\widetilde{\theta}=0.2$.
The accuracy level of our approximations~(\ref{eq:appr_1st})-(\ref{eq:appr_2nd}) for the Bernoulli-Gaussian example is shown in Table~\ref{table:BG}.

It can be seen from Tables~\ref{table:Bernoulli}~and~\ref{table:BG} 
that when the mismatch is minor, the first order approximation can 
achieve relative error below 1\%. However, as the mismatch increases, 
we may need to include the second order term to reduce error. 

\begin{table}[t!]
\caption{Relative error in the Bernoulli-Gaussian example} 
\centering 
\begin{tabular}{c c c c c} 
\hline 
$\widetilde{\theta}$ & $\Delta$ & 1st (\ref{eq:appr_1st}) &2nd (\ref{eq:appr_2nd_appr}) &2nd (\ref{eq:appr_2nd})\\ 
\hline 
0.11 & 0.0006 & 0.25\%  &0.09\%    & 0.09\% \\
0.13 & 0.0052 & 1.4\%   &0.04\%    & 0.08\%\\ 
0.15 & 0.0132 & 3.5\%   &0.21\%    & 0.03\%\\ 
0.17 & 0.0240 & 6.5\%   &0.98\%    & 0.08\%\\ 
0.20 & 0.0444 & 13\%    &4\%    & 0.4\%\\ 
\hline 
\end{tabular}
\label{table:BG} 
\end{table}

\subsection{Amplification for non-i.i.d. signals}

To show that the results of this paper are truly useful,  
it would be interesting to evaluate whether the amplification effect
can be used to characterize the performance of more 
complicated problems that feature mismatch. By their nature, 
complicated problems may be difficult to characterize in closed form, 
not to mention that the theory underlying them may not be well-understood. 
Therefore, we will pursue a setting where a large linear system
is solved for non-i.i.d. signals. Note that neither state 
evolution~\cite{Bayati2011}
nor the decoupling principle~\cite{GuoVerdu2005} 
has been developed for such non-i.i.d. settings, and so any 
predictive ability would be heuristic. 
We now provide such an example, and demonstrate that we can
predict the MSE of one ad-hoc algorithm from that of another.

 \vspace{0.08in}
{\bf Example 3: Markov-constant input.} 
Our Markov-constant signal is generated by a two-state Markov state machine that
contains states $s_0$ (zero state) and $s_1$ (nonzero state). The signal values in
states $s_0$ and $s_1$ are 0 and 1, respectively.
Our transition probabilities are $p(s_0|s_1)=0.2$ and 
$p(s_1|s_0)=1/45$, which yield 10\% nonzeros in the Markov-constant signal. 
In words, this is a block sparse signal whose entries typically stay ``on" 
with value 1 roughly 5 times, and then remain ``off" for roughly 45 entries.
The optimal PME for this non-i.i.d. signal requires the posterior to be 
conditioned on the entire observed sequence, which seems computationally 
intractable. Instead, we consider two sub-optimal yet practical estimators 
where the conditioning is limited to signal entries within windows of sizes 
two or three. We call these practical approaches Bayesian sliding window denoisers:
\begin{align*}
\widehat{x}_2(i) &=\mathbb{E}\left[ x(i)|(y(i-1),y(i))\right],\\
\widehat{x}_3(i) &=\mathbb{E}\left[ x(i)|(y(i-1),y(i),y(i+1))\right].
\end{align*}
Increasing the window-size would capture more statistical information about the signal, 
resulting in better MSE performance. Unfortuntely, a larger window-size requires more 
computation.

\begin{figure}[t!]
\center
\includegraphics[width=\linewidth]{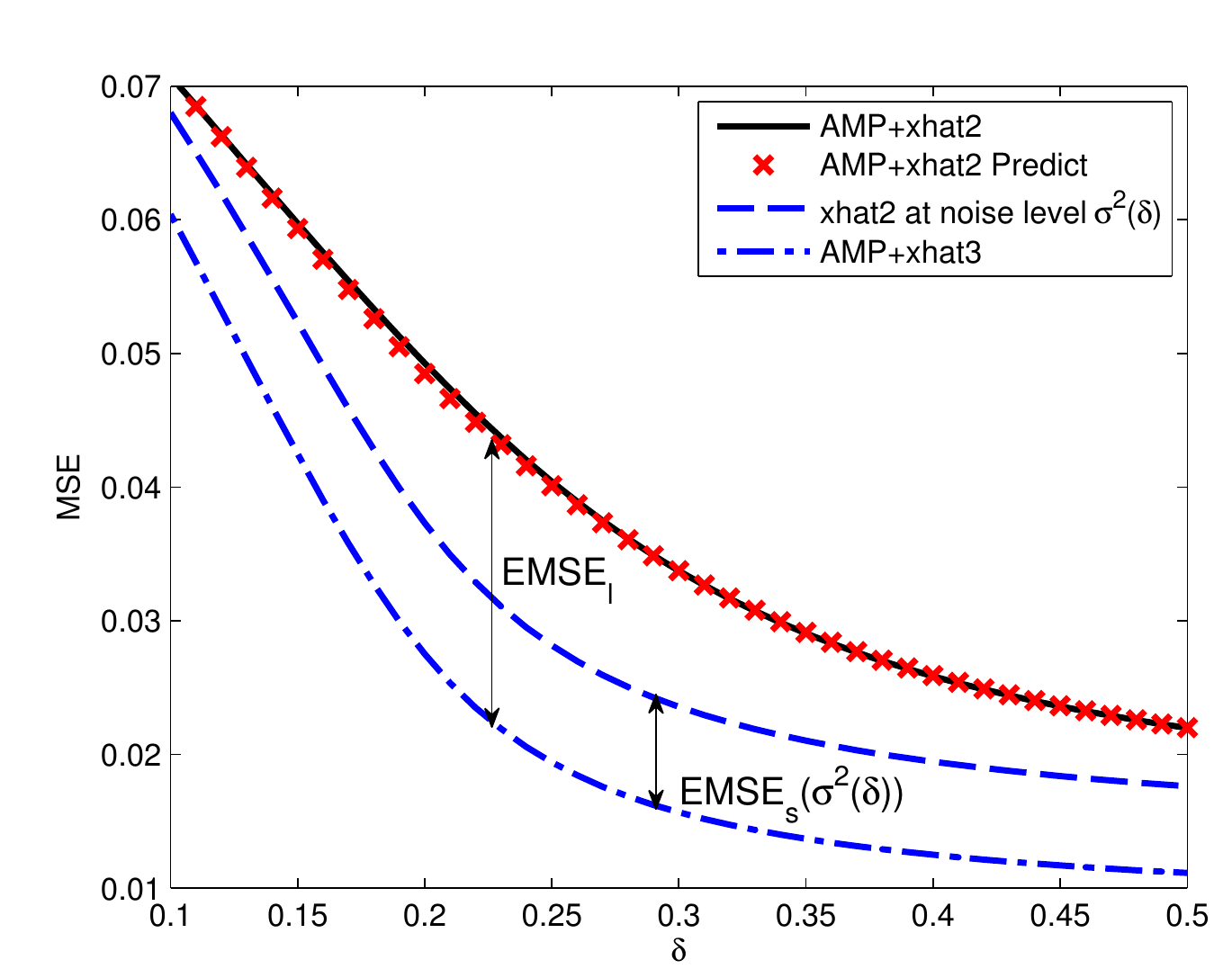}
\caption{Prediction of MSE for Markov-constant input signal 
($p(s_1|s_0)=1/45$ and $p(s_0|s_1)=0.2$) 
in large linear systems. In the legend, 
$\sigma^2(\delta)$ represents the noise variance that AMP with $\widehat{x}_3$ converges to at measurement rate $\delta$. ($\sigma_z^2=0.1$.)}
\label{fig:AmpEffect_Mconst}
\end{figure}

To decide whether a fast and simple yet less accurate 
denoiser should be utilized instead of a more accurate yet slower denoiser
while still achieving an acceptable MSE, we must 
predict the difference in MSE performance between the two.
Such a prediction is illustrated in Figure~\ref{fig:AmpEffect_Mconst}. 
The dash-dotted curve represents the MSE achieved by AMP using 
the better denoiser $\widehat{x}_3$ as a function of the measurement rate $\delta$,
where for each $\delta$, we obtain a solution to the fixed point equation~(\ref{eq:SE}), which is denoted by $\sigma^2(\delta)$. The dashed curve represents the MSE achieved by $\widehat{x}_2$ in scalar channels with noise variance $\sigma^2(\delta)$.
Note, however, that $\widehat{x}_2$ uses window-size 2 whereas $\widehat{x}_3$
uses window-size 3 and can gather more information about the signal.
Therefore, $\widehat{x}_2$ obtains higher (worse) MSE despite denoising a statistically 
identical scalar channel.
The vertical distance between the dashed and dash-dotted  
curves is $\text{EMSE}_s$. 

What MSE performance can be obtained by using the faster denoiser $\widehat{x}_2$
within AMP? The predicted MSE performance using $\widehat{x}_2$ within AMP
is depicted with crosses; the prediction relies on our second order 
approximation~(\ref{eq:appr_2nd}). 
That is, the vertical distance between the crosses and dash-dotted curve is 
$\text{EMSE}_l$~(\ref{eq:appr_2nd}). 
Finally, the solid curve is the true MSE achieved by AMP with $\widehat{x}_2$. 
The reader can see that the predicted MSE performance may help the user 
decide which denoiser to employ within AMP.

Whether the successful predictive ability 
in this example is a coincidence or perhaps connected to a theory of 
decoupling for non-i.i.d. signals remains to be seen. Nevertheless, this example indicates that the formulation (\ref{eq:claim}) that relates the mismatched estimation in scalar channels to that in large linear systems, as well as its closed form approximations (\ref{eq:appr_1st}), (\ref{eq:appr_2nd_appr}), and (\ref{eq:appr_2nd}) can already be applied to far more complicated systems for practical uses.

\section{Conclusion}
\label{sec:conclusion}

We studied the excess mean square error (EMSE) above the 
minimum mean square error (MMSE) in large linear systems due 
to the mismatched prior for the posterior mean estimator (PME). 
In particular, we derived the relationship between the EMSE in 
large linear systems and that in scalar channels;
three simple approximations to this relationship were provided. 
Numerical examples show that our approximations are accurate, 
indicating that they can be used to predict the EMSE in 
large linear systems due to mismatched estimation.

\section*{Appendix: Second Order Approximation of $\text{EMSE}_l$}
\label{Appendix}
Solving (\ref{eq:solving_for_EMSE_l}) for $\text{EMSE}_l$:
\begin{align*}
\text{EMSE}_l&=\frac{\delta}{\beta}\left((\delta-\alpha)\pm\sqrt{(\delta-\alpha)^2-2\beta\text{EMSE}_s(\sigma_p^2)}\right)\\
&+o\left(\Delta^2\right).
\end{align*}

Recall that $\sigma_p^2$ is the noise variance in the decoupled scalar channel for the MMSE estimator, and $\Psi_p(\sigma_p^2)$ is the MMSE. That is, $\sigma_q^2>\sigma_p^2$ and $\Psi_q(\sigma_p^2)>\Psi_p(\sigma_p^2)$, for all $q_X\neq p_X$. Hence, $\Delta=\sigma_q^2-\sigma_p^2>0$ and $\text{EMSE}_s(\sigma_p^2)=\Psi_q(\sigma_p^2)-\Psi_p(\sigma_p^2)>0$. Under the condition that $\Psi_q(\sigma^2)$ is locally concave around $\sigma_p^2$, the second derivative of $\Psi_q(\sigma^2)$ at $\sigma_p^2$ is negative. That is, $\beta<0$.
Therefore, $(\delta-\alpha)^2-2\beta\text{EMSE}_s(\sigma_p^2)\geq (\delta-\alpha)^2$, and
\begin{align}
\text{EMSE}_l&=\frac{\delta}{\beta}\left((\delta-\alpha)-\sqrt{(\delta-\alpha)^2-2\beta\text{EMSE}_s(\sigma_p^2)}\right)\nonumber\\
&+o\left(\Delta^2\right)\nonumber\\
&=\frac{\delta\cdot(\delta-\alpha)}{\beta}\left( 1-\sqrt{1-\frac{2\beta\text{EMSE}_s(\sigma_p^2)}{(\delta-\alpha)^2}}\right)\nonumber\\
&+o\left(\Delta^2\right).\label{eq:appr_2nd}
\end{align}
A Taylor expansion of $\sqrt{1+x}$ yields
\begin{equation*}
\sqrt{1+x} = 1+\frac{1}{2}x-\frac{1}{8}x^2+o(x^2).
\end{equation*}
Let $x=-\frac{2\beta\text{EMSE}_s(\sigma_p^2)}{(\delta-\alpha)^2}$, then
\begin{align}
&\sqrt{1-\frac{2\beta\text{EMSE}_s(\sigma_p^2)}{(\delta-\alpha)^2}}\nonumber\\
&=1-\frac{\beta\text{EMSE}_s(\sigma_p^2)}{(\delta-\alpha)^2}-\frac{1}{2}\frac{\left(\beta\text{EMSE}_s(\sigma_p^2)\right)^2}{(\delta-\alpha)^4}\nonumber\\
&+o\left(\text{EMSE}_s(\sigma_p^2)^2\right).\label{eq:sqrt_x}
\end{align}
Plugging (\ref{eq:sqrt_x}) into~(\ref{eq:appr_2nd}),
\begin{align*}
\text{EMSE}_l 
&=\frac{\delta\text{EMSE}_s(\sigma_p^2)}{(\delta-\alpha)}\left(1+\frac{1}{2}\frac{\beta\text{EMSE}_s(\sigma_p^2))}{(\delta-\alpha)^2}\right)\\
&+o\left(\text{EMSE}_s(\sigma_p^2)^2\right)+o\left(\Delta^2\right).
\end{align*}
Note that
\begin{equation*}
\frac{\text{EMSE}_s(\sigma_p^2)^2}{\Delta^2}\leq\frac{\text{EMSE}_l^2}{\Delta^2}=\delta^2,
\end{equation*}
and so,
\begin{equation*}
\displaystyle\lim_{\Delta\rightarrow 0}\frac{\text{EMSE}_s(\sigma_p^2)^2}{\Delta^2}\leq\delta^2<\infty,
\end{equation*} 
which implies 
that writing $o\left(\text{EMSE}_s(\sigma_p^2)^2\right)+o\left(\Delta^2\right)$ is equivalent to writing $o\left(\Delta^2\right)$.

We have proved the desired result~(\ref{eq:appr_2nd_appr}).

\section*{Acknowledgements}
We thank Dongning Guo, Neri Merhav, Nikhil Krishnan,
and Jin Tan for informative discussions.

\bibliographystyle{IEEEtran}
\bibliography{cites}

\end{document}